# Reversal mode instability and magnetoresistance in perpendicular (Co/Pd)/Cu/(Co/Ni) pseudo-spin-valves

J. E. Davies[1,*], D. A. Gilbert[2], S. M. Mohseni[3,4], R. K. Dumas[5], J. Åkerman[3,4,5] and Kai Liu[2,*]

[1]NVE Corporation, Eden Prairie, MN 55344
[2]Physics Department, University of California, Davis, CA 95616
[3]Materials Physics, School of ICT, Royal Institute of Technology (KTH), 164 40 Kista, Sweden
[4]NanOSC AB, Electrum 205, 164 40 Kista, Sweden
[5]Department of Physics, University of Gothenburg, 412 96 Gothenburg, Sweden

## Abstract

We have observed distinct temperature-dependent magnetization reversal modes in a perpendicular $(Co/Pd)_4/Co/Cu/(Co/Ni)_4/Co$ pseudo-spin-valve, which are correlated with spin-transport properties. At 300 K, magnetization reversal occurs by vertically correlated domains. Below 200 K the hysteresis loop becomes bifurcated due to laterally correlated reversal of the individual stacks. The magnetic configuration change also leads to higher spin disorders and a significant increase in the giant magnetoresistance effect. First order reversal curve measurements reveal that the coupled state can be re-established through field cycling, and allow direct determination of the interlayer coupling strength as a function of temperature.

Giant magnetoresistance (GMR) and tunneling magnetoresistance (TMR)-based magnetic heterostructures with perpendicular magnetic anisotropy (PMA) attract ever increasing attention due to their potential use in spin transfer torque (STT) and other spintronic applications.[1–5] A main challenge in developing these devices with PMA materials is that the optimal non-magnetic spacer thickness for maximizing the MR is also in a regime where strong magnetic coupling exists between the ferromagnetic layers at room temperature.[6,7] For GMR pseudo-spin-valves (PSVs), thickening the spacer helps to decouple the layers,[8] but is often at the expense of reducing the GMR.[9,10] The situation is more extreme in TMR structures as the tunnel barrier thickness dependence is very sensitive.[11,12] In the all-important thin spacer regime, the effect of strong interlayer exchange coupling is usually discussed in terms of the degree of antiferromagnetic coupling across the spacer in zero field, which corresponds to the spin disorder that gives rise to the high resistance state.[13,14]

Here we report a strongly temperature-dependent GMR effect in a (Co/Pd)/ Cu/(Co/Ni) GMR PSV with a thin Cu spacer (30 Å) and the Co/Pd and Co/Ni ML stacks have two very different coercivities at all temperatures. Using first order reversal curve (FORC) measurements we have determined that the GMR temperature-dependence is caused by the change from vertically to laterally (i.e. layer-by-layer) correlated reversal with decreasing temperature. Furthermore, we have directly measured the variation of the interlayer exchange coupling strength.

The PSV in this study was grown by dc-magnetron sputtering onto silicon wafers with a 2000 Å vapor deposited $SiN_x$ buffer layer. The vacuum chamber has a base pressure of 6 x $10^{-8}$ torr, and the Ar gas pressure during sputtering was 3 mtorr. We focus on the following polycrystalline film stack (thicknesses in Å):

SiN$_x$/Ta (50)/Cu (100)/Pd (100)/[Co (4)/Pd (8)]$_4$/Co (4)/Cu (30)/[Co (2)/Ni (4.5)]$_4$/Co (2)/Pd (25)

The Cu spacer thickness is tuned to generate antiferromagnetic (AF) interlayer exchange coupling.[15,20] Magnetic properties were studied by a superconducting quantum interference device (SQUID) and a vibrating sample magnetometer (VSM), with applied field normal to the film plane. Four-point magneto-transport measurements were performed on the VSM. In order to study the magnetization reversal behavior, we have employed a FORC technique, as described earlier.[15–18] After positive saturation, the sample is brought to a reversal field $H_R$, and the magnetization $M(H, H_R)$ is measured as the applied field $H$ increases to positive saturation, tracing out a FORC. A family of FORCs is measured at different $H_R$ to extract a FORC distribution, $\rho(H, H_R) \equiv -\frac{1}{2}\frac{\partial^2 M(H, H_R)}{\partial H \partial H_R}$, which captures *irreversible* switching processes.

The major hysteresis loops and GMR at various temperatures are shown in Figs. 1a and 1b, respectively. At 300 K, along the descending field branch of the major loop, there is a single abrupt magnetization switching event, corresponding to a nucleation field ($H_{N1}$) of 0.40 kOe. $H_{N1}$ decreases monotonically with decreasing temperature. The observation of a single switching event indicates that both the Co/Ni and Co/Pd stacks are switching simultaneously by a vertically correlated reversal mechanism. Below T = 200 K (Fig. 1a) a second switching "corner" emerges at a field ($H_{N2}$) of -0.5 kOe, resulting in bifurcation of the major loop. A significant decrease in $H_{N1}$ is also observed. This bifurcation is due to the decoupling of the Co/Ni and Co/Pd stacks (i.e. laterally correlated reversal). The decoupling becomes more pronounced with decreasing temperature as $H_{N2}$ decreases monotonically to -0.9 kOe at 50 K. A measurement at T = 15 K (not shown) shows that $H_{N2}$ does not substantially decrease below -0.9 kOe. Major loops for individually grown $(Co/Ni)_4/Co$ and $(Co/Pd)_4/Co$ MLs (Fig. 1a, inset) show the Co/Ni ML to be magnetically softer and thus always reverses first in the PSV. It is also worth noting that the individual loops were measured at room temperature and thus the 300 K PSV loop is obviously not a simple superposition of loops from the two layers.

The change in reversal behavior with temperature also has a pronounced effect on the GMR (Fig. 1b). At T = 300 K, a 0.3 % GMR is observed, with a gradually varying dependence on magnetic field. The shape of the GMR curve changes significantly at T = 200 K, with a sharp peak emerging at $H = \pm 0.5$ kOe along the ascending/descending field sweep. As T is further decreased, the peak center shifts to H = -0.7 kOe and GMR increases substantially, reaching 1.6 % at 77 K. The significant GMR enhancement is due to the higher spin disorder associated with the largely antiparallel alignment of the Co/Ni and Co/Pd stacks. The GMR peaks near $H_{N2}$ from the major loop (Fig. 1, vertical dashed line), where the Co/Ni stack is near negative saturation and the Co/Pd stack is just starting to reverse, and the net magnetization is close to zero.

In order to more closely compare the reversal behavior between the vertically and laterally correlated modes we performed FORC measurements at various temperatures. Fig. 2 shows the FORC curves (top) and contour plots of ρ (bottom), whose horizontal axes are aligned to better correlate the key features, at 300 K (left) and 77 K (right). The starting points of the FORCs delineate the descending branch of the major loop. SQUID-VSM measurements (not shown) show the ascending branch to be symmetric. At 300K, the major loop is pinched near the coercive field (Fig. 2a). The FORC distribution (Fig. 2c) exhibits a horizontal ridge, a plateau and a negative/positive region as $H_R$ is decreased (denoted by areas 1, 2 and 3, respectively).

This pattern is characteristic of vertically correlated reversal by a domain nucleation, propagation and annihilation process, as previously seen in Co/Pt and (Co/Pt)/Ru systems.[17,19,20] This is also supported by a magnetic force microscopy (MFM) study after ac-demagnetization (Fig. 2a, inset) showing a labyrinth-type domain pattern with domain width of ~100 nm.

The FORCs measured at 77 K clearly show the major loop bifurcation (i.e. the laterally correlated reversal mode) with the second switching "corner" forming at $H_{N2}$ = -0.9 kOe (Fig. 2b) along the descending branch. With the bifurcation, the FORCs change significantly with a noticeable grouping around $H$=0.5 kOe (Fig. 2b, circled region) as the minor loop from the decoupled magnetization reversal of the Co/Ni ML develops. The change in reversal behavior is accompanied by a more complex FORC distribution (Fig. 2d), characteristic of the laterally correlated mode. As $H_R$ is decreased from positive saturation the first feature encountered is a positive peak at $H_R$ = -0.25 kOe (Fig. 2d, feature 1), due to the Co/Ni reversal. For -0.9 kOe < $H_R$ < -0.8 kOe, coincident with the onset of Co/Pd reversal, a negative "pit" emerges in the FORC distribution, followed by a positive horizontal ridge (Fig. 2d, feature 2). In this $H_R$ range, the slope of the successive FORCs decreases near $H$ = 0.5 kOe (Fig. 2b, circled region), leading to the negative feature, but increases at higher $H$-values, resulting in the positive ridge. This is an indication of a significant change to the vertically correlated reversal mode. Finally, for $H_R$ < -0.9 kOe $\rho$ becomes a complicated series of two negative/positive features (Fig. 2d, region 3). These features arise as the vertically correlated mode reverts to the laterally correlated mode. This is also seen in the FORCs (Fig. 2b, dashed boxed region), and the FORCs protrude outside of the major loop, a direct consequence of the reversal mode instability.

To better illustrate the crossover between reversal modes, Fig. 3 shows several representative FORCs at 77 K. The Co/Ni ML reverses at $H_{N2} < H_R < H_{N1}$. At $H_R$ = -0.6 kOe the Co/Ni ML is near negative saturation; the FORC shows an abrupt switching at $H$ = 0.35 kOe, corresponding to re-nucleation of reversal domains in the positive field direction (filled squares). For $H_R < H_{N2}$ the Co/Pd ML undergoes reversal. At $H_R$ = -1.1 kOe domains are present in the Co/Pd ML while the Co/Ni ML is saturated. The FORC curvature here has changed significantly, with the reversal becoming more gradual and no distinct renucleation steps are present (open squares). When FORCs measured at 300 K at similar $H_R$ are superimposed onto Fig. 3 (dashed lines), the resemblance in both curvature and saturation field to the 77 K FORCs is striking. This is a clear indication that the vertically correlated mode is reestablished by field cycling. It is only near complete saturation at $H_R$ = -3.0 kOe, well beyond the apparent saturation of the major loop at -2.1 kOe, that the FORCs return to the bifurcated and stepped reversal exhibited by the descending branch of the major loop (filled triangles).

This exotic crossover between vertically and laterally correlated reversal modes with both temperature and field cycling highlights the delicate balance amongst the magnetostatic ($E_{demag}$), domain wall ($E_{wall}$) energies and interlayer exchange coupling ($E_{ex}$), which is typically AF at our Cu thickness.[15,13,20] The stable state is defined by the difference in total energy between the laterally and vertically correlated modes. The saturation magnetization ($M_s$) and the perpendicular anisotropy ($K_u$), the parameters which define $E_{demag}$ and $E_{wall}$, have a small effect on the energy difference. $M_s$ increases by 5 % between 300 K and 77 K based on our measurements and $E_{wall}$ is proportional to $\sqrt{K_U}$ resulting roughly 5 % variation as well based on literature values.[21] Thus, the dominating parameter controlling the reversal process is $E_{ex}$, and can be quantified by the exchange energy density, $J=H_E M_s t$, where $H_E$ and $t$ are the exchange field and thickness of the switching magnetic layer, respectively. Values of $M_S$ = 775 emu/cm$^3$ and $t$ = 2.8 nm were used based on the measurement of the Co/Ni ML grown individually (Fig. 1a, inset). We are able to extract $H_E$ of the Co/Ni ML for T ≤ 150 K from its peak position in $\rho$ (Fig. 2d, feature 1), using a coordinate transformation where the local bias field is given by $H_b = (H + H_R)/2$.[22,23] Values of $H_E$ and the corresponding $J$ are shown in Table I, which increase by a factor of ~ 2.5 from 150 K to 77K.[24] It is this strong temperature dependence that tips the energy balance and leads to the reversal mode crossover. These values of $J$ are the same order of magnitude as those reported for Co/Cu/Co trilayers.[15]

We also observe that slight changes to the ML stack structure can destroy the ability to switch between the modes. This is evidenced in a similar PSV with the only change being the addition of a Ni/Co bilayer at the Cu spacer on the Co/Pd ML side:

SiNx/Ta(50)/Cu(100)/Pd(100)/[Co(4)/Pd(8)]$_4$/**Ni(5)/Co(4)**/Cu(30)/[Co(2)/Ni(4.5)]$_4$/Co(2)/Pd(25)

The major loops for this sample (Fig. 4a) show no evidence of bifurcation and maintain the pinched shape, indicating persistence of the vertically correlated reversal mode at all temperatures. An MFM micrograph (Fig. 4a, inset) shows that the domain topography and domain size is similar to the earlier sample (Fig. 2a inset). Additionally, the GMR curves (Fig. 4b) remain qualitatively similar with no sharp increase in GMR below 200 K as was observed in our mixed mode sample (Fig. 1b). The maximum resistance occurs near the coercive field where the labyrinth domain pattern results in the largest spin disorder. The inability of this sample to achieve the laterally correlated reversal mode is likely due to an increase in magnetostatic energy ($\propto M_s^2$) in the Co/Pd stack from the additional moment as well as an anisotropy reduction from the softer Ni layer reducing the domain wall energy ($\propto \sqrt{K_U}$), making domain nucleation in that stack more favorable. This would also increase the amount of dipole field that would be imparted on the Co/Ni stack from the demagnetized Co/Pd ML, resulting in reinforcement of the vertical correlation of domains even at decreased temperatures.

Interestingly, such a delicate balance has been previously observed in more strongly AF coupled Co/Pt MLs with Ru spacers, where temperature and field cycling were able to induce the laterally and vertically correlated reversal modes. In that case, the exchange coupling was more pronounced with *J* being an order of magnitude larger than the present system, consistent with the more pronounced AF coupling across the Ru spacers.[20,24]

In conclusion, we have demonstrated the crossover between vertically and laterally correlated magnetization reversal in a (Co/Pd)/Cu/(Co/Ni) PSV, which can be manipulated by both temperature variation and cycling of the magnetic field. The contrasting magnetic configurations offer a new handle to tune spin disorders and GMR effects. The instability of these modes is due to the delicate balance amongst magnetostatic, domain wall and interlayer exchange coupling energies. The FORC measurements were utilized to extract the exchange energy density, which has strong temperature dependence that facilies the reversal mode crossover. Adding an additional Ni/Co bilayer at the [Co/Pd]/Cu interface further demonstrates the sensitivity of the mixed reversal behavior and was able to completely suppress the laterally coupled reversal mode, likely due to increased $M_s$ and reduced $K_u$.

The authors would like to acknowledge support from the National Science Foundation (Grant # IIP-1013982, DMR-1008791 and ECCS-0925626), the Swedish Institute, the Swedish

Foundation for Strategic Research, the Sweidsh Research Council, and the Knut and Alice Wallenberg Foundation. JED would like to acknowledge Dave Brownell for useful discussions.

# References

*Authors to whom correspondence should be addressed. Electronic mail: jdavies@nve.com, kailiu@ucdavis.edu.

[1] S. Mangin, D. Ravelosona, J.A. Katine, M.J. Carey, B.D. Terris, and E.E. Fullerton, Nature Materials **5**, 210 (2006).

[2] D.C. Ralph and M.D. Stiles, Journal of Magnetism and Magnetic Materials **320**, 1190 (2008).

[3] J.A. Katine and E.E. Fullerton, Journal of Magnetism and Magnetic Materials **320**, 1217 (2008).

[4] W.H. Rippard, A.M. Deac, M.R. Pufall, J.M. Shaw, M.W. Keller, S.E. Russek, and C. Serpico, Physical Review B **81**, 014426 (2010).

[5] S.A. Wolf, D.D. Awschalom, R.A. Buhrman, J.M. Daughton, S. von Molnár, M.L. Roukes, A.Y. Chtchelkanova, and D.M. Treger, Science (New York, N.Y.) **294**, 1488 (2001).

[6] C.L. Zha, Y.Y. Fang, J. Nogués, and J. Åkerman, Journal of Applied Physics **106**, 053909 (2009).

[7] G. Feng, H.C. Wu, J.F. Feng, and J.M.D. Coey, Applied Physics Letters **99**, 042502 (2011).

[8] S. Mohseni, R. Dumas, Y. Fang, J. Lau, S. Sani, J. Persson, and J. Åkerman, Physical Review B **84**, 174432 (2011).

[9] M.N. Baibich, J.M. Broto, A. Fert, F.N. Van Dau, and F. Petroff, Physical Review Letters **61**, 2472 (1988).

[10] B. Dieny, V. Speriosu, S. Parkin, B. Gurney, D. Wilhoit, and D. Mauri, Physical Review B **43**, 1297 (1991).

[11] J.S. Moodera and G. Mathon, Journal of Magnetism and Magnetic Materials **200**, 248 (1999).

[12] S.S.P. Parkin, K.P. Roche, M.G. Samant, P.M. Rice, R.B. Beyers, R.E. Scheuerlein, E.J. O'Sullivan, S.L. Brown, J. Bucchigano, D.W. Abraham, Y. Lu, M. Rooks, P.L. Trouilloud, R.A. Wanner, and W.J. Gallagher, Journal of Applied Physics **85**, 5828 (1999).

[13] S. Parkin, R. Bhadra, and K. Roche, Physical Review Letters **66**, 2152 (1991).

[14] J. Xiao, J. Jiang, and C. Chien, Physical Review Letters **68**, 3749 (1992).

[15] I. Mayergoyz, *Mathematical Models of Hysteresis and Their Applications* (AcademicPress-Elsevier, 2003).

[16] A.P. Roberts, C.R. Pike, and K.L. Verosub, Journal of Geophysical Research **105**, 28461 (2000).

[17] J. Davies, O. Hellwig, E. Fullerton, G. Denbeaux, J. Kortright, and K. Liu, Physical Review B **70**, 224434 (2004).

[18] J.E. Davies, O. Hellwig, E.E. Fullerton, J.S. Jiang, S.D. Bader, G.T. Zimányi, and K. Liu, Applied Physics Letters **86**, 262503 (2005).

[19] M. Pierce, C. Buechler, L. Sorensen, J. Turner, S. Kevan, E. Jagla, J. Deutsch, T. Mai, O. Narayan, J. Davies, K. Liu, J. Dunn, K. Chesnel, J. Kortright, O. Hellwig, and E. Fullerton., Physical Review Letters **94**, 17202 (2005).

[20] O. Hellwig, A. Berger, J.B. Kortright, and E.E. Fullerton, Journal of Magnetism and Magnetic Materials **319**, 13 (2007).

[21] J.I. Hong, S. Sankar, A.E. Berkowitz, and W.F. Egelhoff, Journal of Magnetism and Magnetic Materials **285**, 359 (2005).

[22] J. Davies, J. Wu, C. Leighton, and K. Liu, Physical Review B **72**, 134419 (2005).

[23] R. Dumas, C.-P. Li, I. Roshchin, I. Schuller, and K. Liu, Physical Review B **75**, 134405 (2007).

[24] J. Davies, O. Hellwig, E. Fullerton, and K. Liu, Physical Review B **77**, 014421 (2008).

[25] O. Hellwig, T.L. Kirk, J.B. Kortright, A. Berger, and E.E. Fullerton, Nature Materials **2**, 112 (2003).

Table I. $H_E$ and $J$ as a function of temperature.

| Temperature (K) | $H_E$ (kOe) | $J$ ($\times 10^{-2}$ erg/cm$^2$) |
|---|---|---|
| 77  | 0.13 | 2.8 |
| 100 | 0.12 | 2.4 |
| 125 | 0.11 | 2.3 |
| 150 | 0.05 | 1.1 |

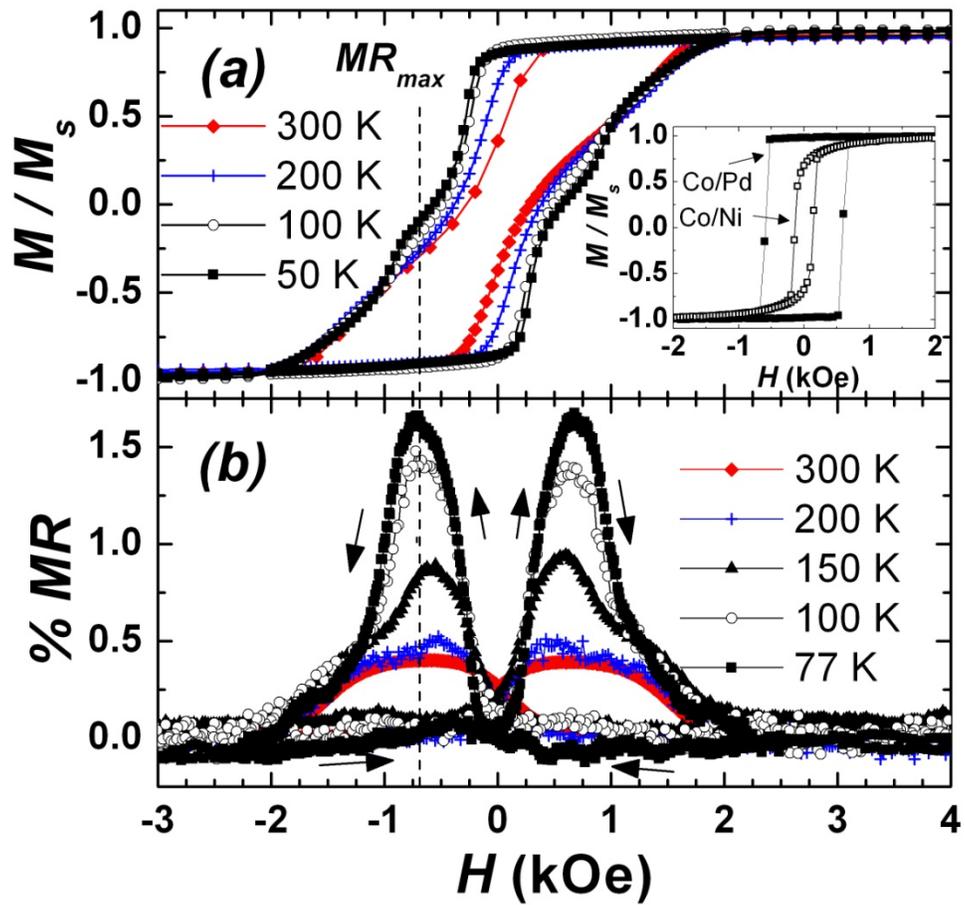

Figure 1 (color online) (a) Major hysteresis loops measured by SQUID Magnetometry at different temperatures, with 300K VSM measurements of the individual Co/Ni and Co/Pd stacks shown in the inset, and corresponding (b) GMR of the (Co/Pd)/Cu/(Co/Ni) PSV. The vertical dashed line highlights the GMR peak position below 100 K. For both measurements the magnetic field is applied perpendicular to the film.

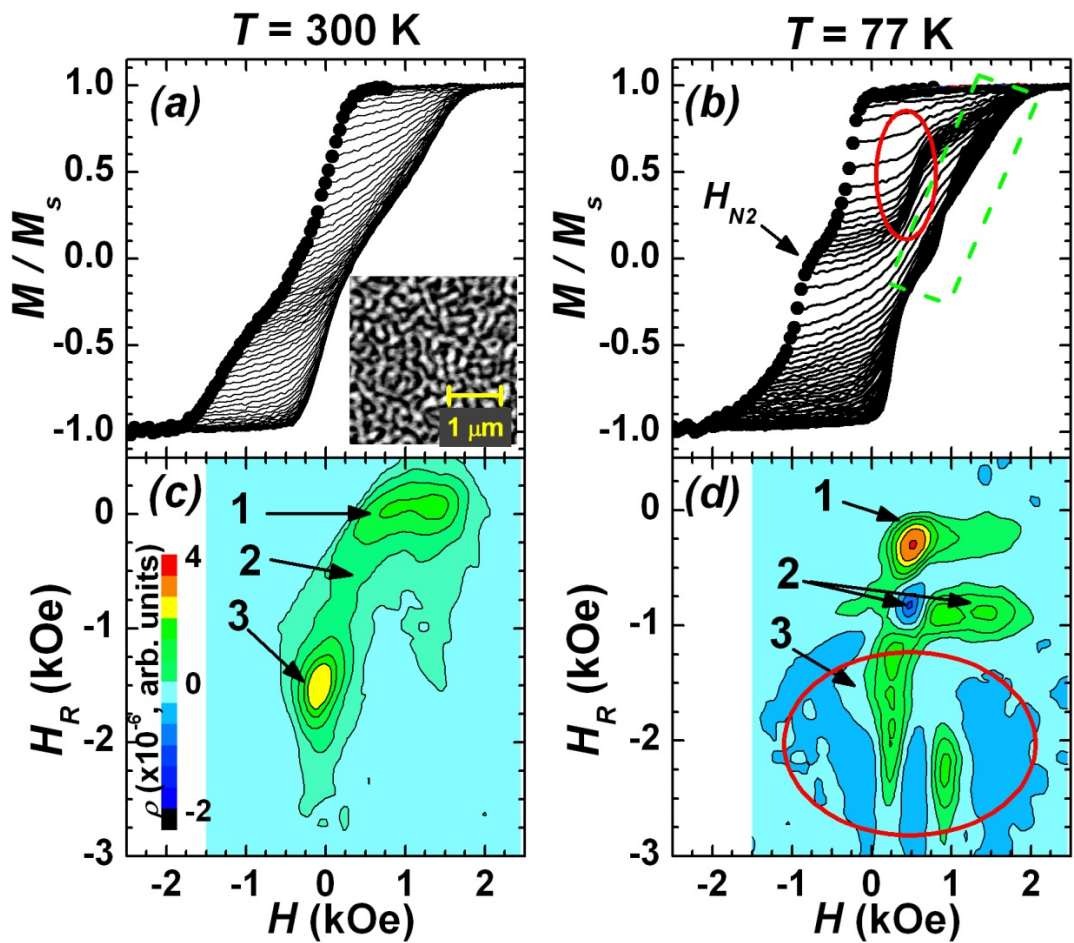

Figure 2 (color online) (Top) FORCs and (bottom) contour plots of ρ for the (Co/Pd)/Cu/(Co/Ni) PSV for (a)/(c) 300 K, (b)/(d) 77 K. The horizontal axes of the contour plots are aligned with those of the FORCs and the contours are normalized to the same scale shown in panel (c). Inset in (a) shows a MFM image at 300 K after ac demagnetization.

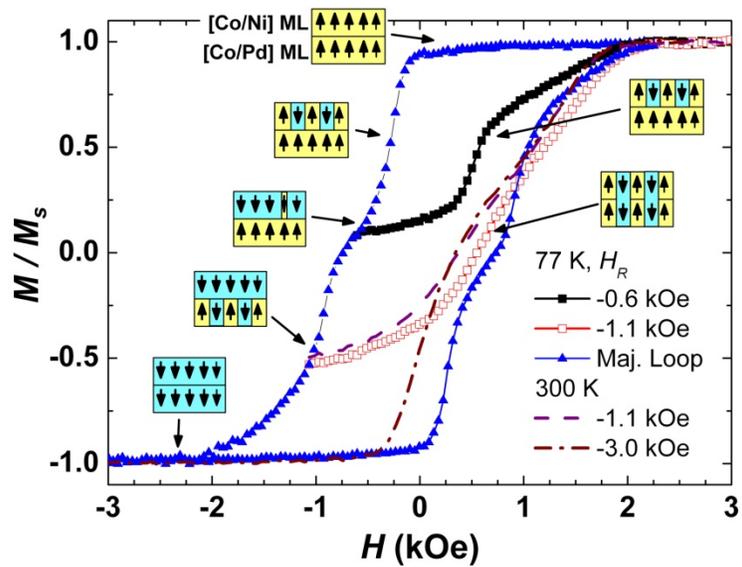

Figure 3. (color online) Major loop (blue triangles) along with representative FORCs and schematics of the reversal behavior at 77K. The -0.6 kOe FORC (filled squares) has a minor loop switching "corner" corresponding to Co/Ni reversal. The -1.1 kOe FORC at 77K (open squares) is similar to the 300 K FORC (dashed line), indicating a return to vertically correlated reversal. Laterally correlated reversal returns near negative saturation (blue triangles).

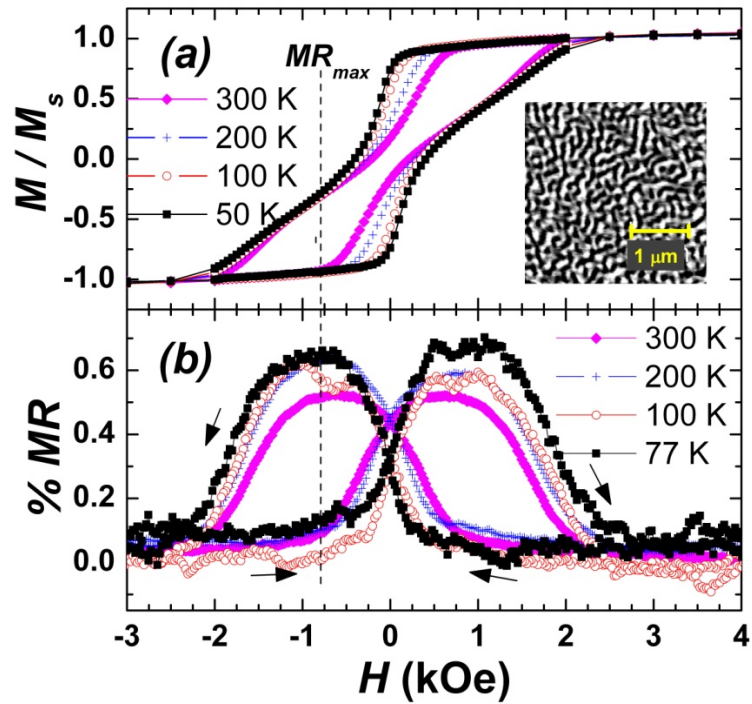

Figure 4. (color online) (a) Major loops and (b) GMR curves at different temperatures for a $(Co/Pd)_4/Ni/Co/Cu/(Co/Ni)_4$ PSV. Inset in (a) is a MFM micrograph at 300 K after ac demagnetization.